\documentclass[11pt]{article}
\usepackage{hyperref}
\pdfoutput=1
\usepackage{hyperref}
\begin{document}

\title{The birth, life and death of convective plumes \\ generated by a green laser in an isothermal fluid}

{\author{Albert N. Sharifulin  \\
\\\vspace{6pt} Department of Applied Physics, \\ Perm State Technical University, Perm, 614990,
Russia
\\\\ Anatoly N. Poludnitsin  \\
\\\vspace{6pt}
Department of Common Physics, \\ Perm State University, Perm,
614000, Russia}} \maketitle
%%  The abstract (in this file, and that submitted as text to arXiv) should
%%  include the exact phrase
%% "fluid dynamics video" or "fluid dynamics videos"
\begin{abstract}
The aim is to laboratory simulation of convective plumes in the
lower mantle, generated by a hot spot on the Earth's core. In the
fluid dynamics video, presented results of experimental modeling
of the plume from the hot point generated by the laser.
Demonstrated that in a fluid with a high Prandtl number, point
heating can generate complex spiral plumes. This experimental
result allows us to question the classical notion of a mantle
plume, as the column of heat (Campbell, IH, 2006, Large Igneous
Provinces and the mantle plume hypothesis, Elements, 1, 265-269,
2006).
\end{abstract}
% main text
\begin{enumerate}
\item Experiment 1. Convective plume generated by
a green laser in an water, Prandtl number Pr=7. Cubic cavity
25x25x25 cm.

\item Experiment 2. Spiral convective plume generated by
 a green laser in an silicon oil with Prandtl number Pr=50. Cubic cavity
9x9x9 cm.

\item Purpose of investigation: Laboratory scale modeling of mantle plumes
generation from hot point on Earth core.

\item Equipment: 0.2W green laser, shlieren dev IAB-451,
digital photo camera Canon EOS 50D.
\end{enumerate}
\end{document}